\documentclass[%
aip,
amsmath,amssymb,
reprint,%
]{revtex4-1}
\makeatletter
\def\@seccntformat#1{\csname the#1\endcsname\quad} 
\makeatother

\usepackage{braket}
\usepackage{graphicx}
\usepackage{graphicx}
\usepackage{mathtools}
\usepackage{amssymb}
\usepackage{amsthm}
\usepackage{amsmath}
\usepackage{subcaption} 
\usepackage{siunitx}
\usepackage{booktabs}
\usepackage{float}
\usepackage{flushend} 
\usepackage[table]{xcolor}\usepackage{placeins}
\usepackage{hyperref}
\begin{document}
	
	\preprint{AIP/123-QED}

\title{Robust controlled-Z gate for Rydberg atoms based on level-crossing-free echoing rapid adiabatic passage}

\author{Yichi Zhang}
\email{zhangyichi@sxu.edu.cn}
\author{Zhenqi Bai}
\author{Xu Zhao}
\author{Hongyan Fan}
\author{Ximo Wang}
\author{Tiecheng Wang}
\affiliation{College of Physics and Electronic Engineering, Shanxi University, 030006 Taiyuan, People’s Republic of China}
\affiliation{Collaborative Innovation Center of Extreme Optics, Shanxi University, Taiyuan, Shanxi 030006, People’s Republic of China}

\date{\today}

\begin{abstract}
We propose a controlled-Z gate scheme for Rydberg atoms based on level-crossing-free echoing rapid adiabatic population transfer. We design antisymmetric Rabi frequency pulses and symmetric detuning pulses, enabling the system to completely avoid level-crossing points throughout the evolution, and the dynamical phase is naturally eliminated by the time-reversal symmetry of the double-pulse sequence. We incorporate dissipative effects through the Lindblad master equation. The numerical simulation yields a two-qubit CZ gate fidelity of 0.9999. When the Rabi-frequency fluctuation is within $\pm 2\%$, and the detuning offset is within $\pm 1\%$, the fidelity can still remain above 0.999. Under the same dissipative model, the three-qubit CCZ gate achieves a fidelity of 0.999. When a single-parameter fluctuation does not exceed $\pm 3\%$, the fidelity is always higher than 0.997. Our scheme requires no laser phase jumps or fast switching operations. The zero-area pulse structure suppresses first-order intensity noise, and the symmetric double-pulse sequence avoids spatially resolved laser switching, making it suitable for parallel gate operations in large-scale neutral-atom arrays.
\end{abstract}

\maketitle

Neutral atom tweezer arrays, with their long coherence times, flexible arrangement, and scalability, have become one of the promising candidate platforms for realizing quantum computing and quantum simulation~\cite{Evered2023, Saffman2010, Shi2022, Bluvstein2022, Scholl2021}. In these systems, strong van der Waals interactions between Rydberg-state atoms prevent simultaneous excitation of neighboring atoms to high-lying states. This phenomenon is known as the Rydberg blockade effect~\cite{Urban2009, Wilk2010, Heidemann2007, Bharti2023}, providing a direct physical mechanism for realizing many-body quantum entanglement and controlled quantum gates~\cite{Saffman2005, Liang2025, Ma2023}. The two-qubit gate fidelities achieved with neutral atoms still lag behind those demonstrated in superconducting systems~\cite{Barends2014, Linke2017, Setiawan2023, Li2023b, Kjaergaard2020, Arute2019, Wu2021b} and ion traps~\cite{Ballance2016, Leu2023, Shapira2023}. In recent years, experiments based on alkali-metal atoms have made significant progress in single-qubit and two-qubit gate operations, with CZ gate fidelity exceeding 0.995~\cite{Levine2019, Graham2019, Wang2026, Isenhower2010, Gaetan2009, Theis2016, Pagano2022, Chang2023, Sung2021, Fu2022}. Very recent experiments in optical tweezers and optical clocks have further extended the capability of neutral-atom multi-qubit gates~\cite{Finkelstein2024, Cao2024}.

Rapid adiabatic passage (RAP) transfer is a widely applied quantum state manipulation technique. Conventional schemes implement frequency chirping to sweep the detuning through resonance, combined with Gaussian Rabi frequency pulses to achieve atomic population inversion~\cite{Shore2017}. However, at the level crossing point, the instantaneous eigenenergy gap of the system is extremely small, making nonadiabatic transitions highly probable and thereby limiting the theoretical ceiling of gate operation fidelity. To overcome this difficulty, various shortcut-to-adiabaticity techniques have been developed~\cite{Petrosyan2017, Jandura2022, Moller2008, Goerz2017, GueryOdelin2019, Kang2022, Huang2018, Zhang2012, Hou2024}, but such methods often require complex time-dependent control of the Hamiltonian, with high experimental implementation thresholds.

Another approach is to alter the pulse symmetry. If the Rabi frequency is a time-odd function and the detuning is a time-even function, the system can complete full population return without level crossing~\cite{Rangelov2010}. This level-crossing-free scheme significantly reduces the risk of nonadiabatic leakage, and the zero-area pulse condition naturally suppresses certain systematic errors. STIRAP technology, as an important extension of RAP, has been systematically developed in physics, chemistry, and broader fields~\cite{Vitanov2017, Boradjiev2010, Alarcon2026}. In the field of Rydberg quantum gates, studies such as adiabatic passage schemes based on Stark-tuned F\"orster resonance~\cite{Beterov2016} and fast Rydberg antiblockade gates~\cite{Su2017, Song2024, Jin2024, Xiao2024, Xu2012, Shi2017, Liu2020} have demonstrated diverse quantum state manipulation pathways.

In the field of Rydberg quantum gates, symmetric double-pulse driving has been proven to be an effective way to eliminate dynamical phases while retaining geometric phases. In recent work, a controlled-Z gate scheme based on two identical RAP pulses was developed; it was proved that two-pulse symmetric driving can effectively cancel dynamical phases while accumulating geometric phases to realize the CZ gate~\cite{Xue2024, Xu2024}. However, the echo method can effectively cancel the dynamical phase while accumulating geometric phase to implement the CZ gate and suppress decoherence over longer gate times, which makes the echo method widely used in quantum gates~\cite{Nguyen2025, Zhou2025, Muniz2025}. Nevertheless, existing echo methods are typically built upon traditional adiabatic evolution and still retain the inherent level-crossing structure~\cite{Vitanov2017, Fang2022, Beterov2026}. At the crossing points, the instantaneous eigenenergy gap approaches zero, making the system highly susceptible to nonadiabatic transitions, thereby limiting the theoretical upper bound of gate operation fidelity~\cite{Liang2025b, Vaezi2022, Liu2020b, Lv2020, Xu2022}.
\begin{figure}[htbp]
	\includegraphics[width=1\linewidth]{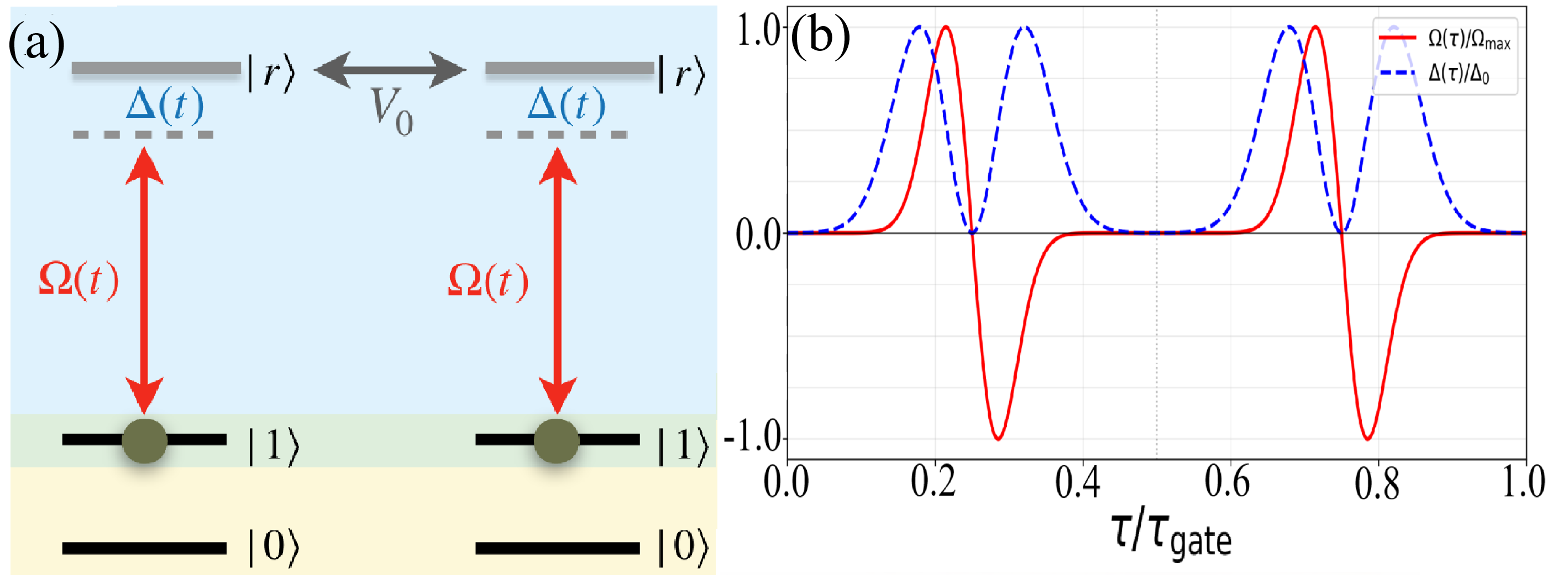}
		\captionsetup{justification=raggedright,singlelinecheck=false}
	\caption{Schematic of the pulse design and atomic--evel structure. (a)Atomic level structure of two-atom qubits featuring Rydberg--Rydberg interactions between the $|r\rangle$ states. (b)Two identical level-crossing-free pulses.
		The Rabi frequency $\Omega(t)$ exhibits an antisymmetric bipolar structure, crossing zero and changing sign at the pulse center;
		the detuning $\Delta(t)$ has a single-peak symmetric structure, smoothly approaching zero at the beginning and end.}
	\label{fig:1}
\end{figure}

In this paper, we employ level-crossing-free echo RAP pulses: using antisymmetric Rabi frequency pulses and symmetric detuning pulses to form a two-segment sequence, enabling a single atom to complete full population return, while the two-atom system acquires a geometric phase under Rydberg blockade. We comprehensively analyze the robustness of this scheme to Rabi frequency and detuning fluctuations, verifying its feasibility in multi-qubit extensions. Compared with existing methods, this level-crossing-free design avoids the region of minimum eigenenergy gap by keeping the detuning strictly positive and only touching resonance at the pulse center, thereby suppressing nonadiabatic transitions. The zero-area Rabi frequency pulse enables first-order suppression of laser intensity noise, improving the gate's tolerance to experimental imperfections. The two-segment symmetric structure cancels the dynamical phase via time-reversal symmetry, eliminating the need for additional phase compensation. This scheme does not require reoptimization of the pulse shape when extending the CZ gate to the CCZ gate. All driving fields are smoothly ramped down at the end of the gate operation, and global laser addressing removes the requirement for rapid spatially resolved switching, supporting parallel execution of multi-pair qubit gates in large-scale arrays.

The remainder of this paper is organized as follows. We first introduce the physical model of the level-crossing-free echoing RAP pulse and the theoretical scheme for constructing the CZ gate, and then extend this scheme to the CCZ gate. Subsequently, we evaluate the gate fidelity and its dependence on interaction strength through numerical simulations, and analyze the robustness under parameter fluctuations. Finally, we give our conclusions.

We design the Rabi frequency of level-crossing-free pulses as a time-odd function satisfying the zero-area condition, which provides first-order suppression of laser intensity noise. RAP pulses lack this constraint and typically require the driving field frequency to chirp through the resonance point. Our level-crossing-free scheme employs two completely identical symmetric pulses; the dynamical phase is naturally canceled due to time-reversal symmetry, requiring no additional laser phase jumps. In contrast, traditional RAP schemes necessitate precise phase control or rely on adiabatic following to avoid residual phases.

We first consider a pair of atoms with Rydberg interaction in Fig.~\ref{fig:1}(a), each atom having two hyperfine ground states $|0\rangle$ and $|1\rangle$ and a Rydberg excited state $|r\rangle$. Quantum information is encoded in the ground states $|0\rangle$ and $|1\rangle$. Through a global laser field, $|1\rangle$ and $|r\rangle$ are coupled with Rabi frequency $\Omega(t)$ and detuning $\Delta(t)$. Both atoms are subject to identical laser driving, and the system Hamiltonian can be written as
\begin{equation}\label{eq:H}
\hat{H}(t) = \sum_{i=1}^{2} \left[\frac{\Omega(t)}{2}|r\rangle_i\langle 1|_i + \text{h.c.}\right] + \Delta(t)|r\rangle_i\langle r|_i + H_I,
\end{equation}
here $H_I = V_0 |r\rangle_1\langle r|_1 \otimes |r\rangle_2\langle r|_2$, where $V_0$ is the Rydberg-Rydberg interaction strength. When $V_0$ is much larger than the Rabi frequency, the doubly excited state $|rr\rangle$ is effectively blocked, and the system evolution is restricted to the single-excitation subspace.

For the initial state $|11\rangle$, both atoms can be simultaneously excited to a superposition of single Rydberg states; for $|01\rangle$ or $|10\rangle$, only a single atom participates in the laser coupling; $|00\rangle$ is a dark state, unaffected by the driving.

We evaluate the gate performance using the gate fidelity, defined as
\begin{equation}\label{eq:fidelity}
\mathcal{F} = \langle\psi|U_{CZ}^\dagger \rho_f U_{CZ}|\psi\rangle,
\end{equation}
where $U_{CZ}$ is the unitary operator of the ideal CZ gate, and $\rho_f$ is the density matrix after actual evolution. The initial state is chosen as the symmetric superposition of all computational basis states, $|\psi\rangle = 1/{\sqrt{2^n}} \sum_{i,j,k,\ldots=0}^{1} |ijk\ldots\rangle$, applicable to an $n$-qubit system, which makes the fidelity sensitive to various errors~\cite{Jiang2023, Pelegri2022}. Spin squeezed states, as important quantum resources, and coherent photon manipulation technology~\cite{Thompson2017} provide rich theoretical backgrounds for understanding collective quantum behaviors in atomic arrays~\cite{Magesan2012}.
\begin{figure}[htbp]
	\includegraphics[width=0.5\textwidth]{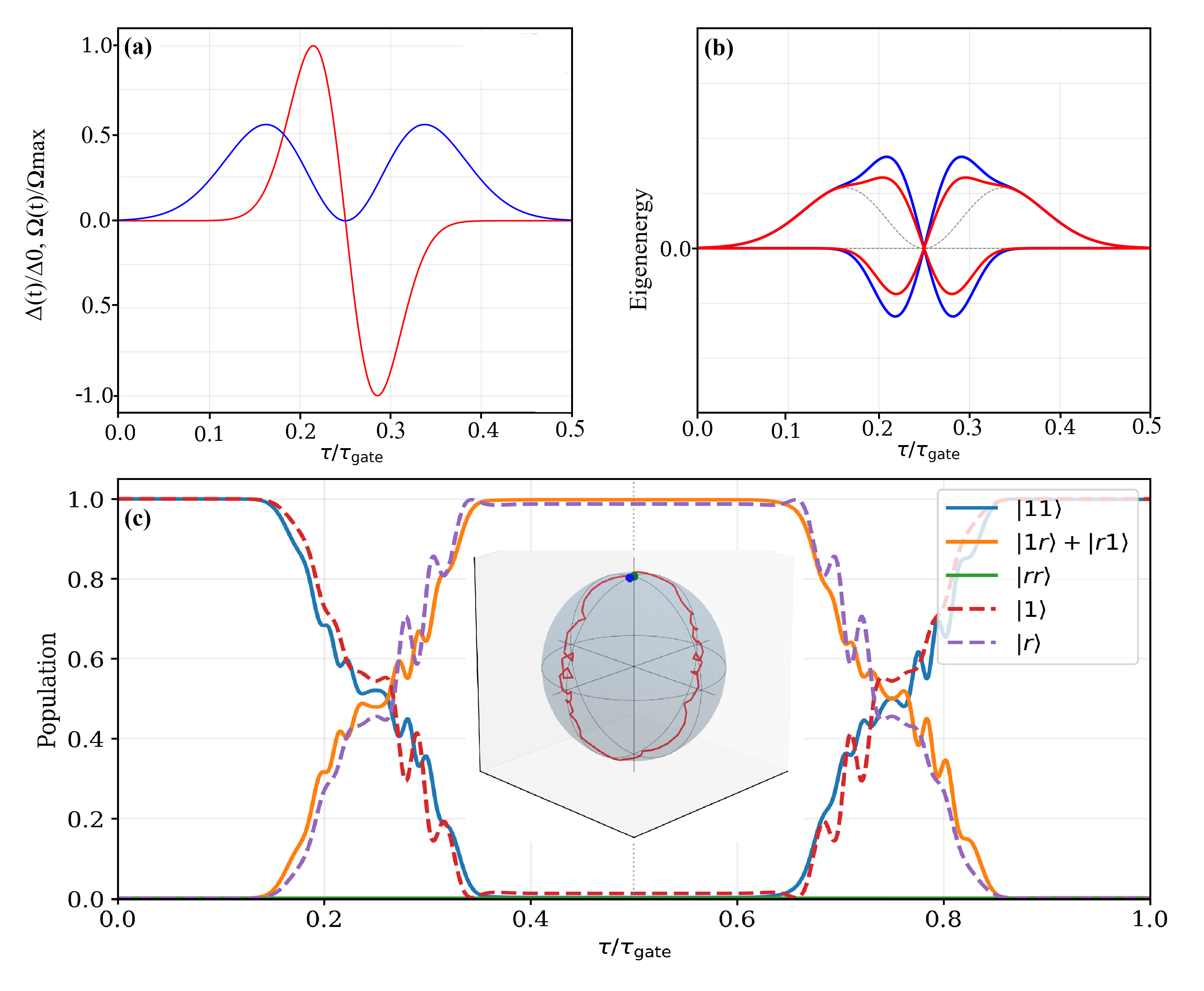}
	\captionsetup{justification=raggedright,singlelinecheck=false}
	\caption{Pulse design and system evolution for CZ and CCZ gates. (a) Profile of the pulses, $\Delta(t)$ and $\Omega(t)$, for CZ and CCZ gate construction. 
		(b) Instantaneous eigenenergies of the Hamiltonian $H(t)$ during one pulse: the blue and red solid lines represent the eigenstates of the two-atom and single-atom systems, respectively. (c) The atomic population evolution during the complete gate time $\tau_{gate}$. The solid line represents the states
		evolution of the double-atom case, and the dashed line represents the single-atom case. Inset: Bloch sphere trajectory
		of the dynamical state. }
	\label{fig:2}
\end{figure}

The pulse profile depicted in Fig.~\ref{fig:1}(b) is intrinsically different from that of conventional RAP protocols. Introducing the dimensionless time $\tau = \Omega_{\text{max}} t$ (with $\Omega_{\text{max}}$ as the maximum Rabi frequency), the Rabi frequency and detuning during the $k$-th ($k = 1, 2$) pulse interval are taken as in~\cite{Rangelov2010}.
\begin{subequations}\label{eq:pulses}
\begin{align}
\Omega_k(\tau) &= -\Omega_{\text{max}} C_\Omega \frac{\tau - \tau_k}{\tau_R} \exp\left[-\left(\frac{\tau - \tau_k}{\tau_R}\right)^2\right], \label{eq:Omega} \\
\Delta_k(\tau) &= \Delta_0 C_\Delta \left(\frac{\tau - \tau_k}{\tau_D}\right)^2 \exp\left[-\left(\frac{\tau - \tau_k}{\tau_D}\right)^2\right], \label{eq:Delta}
\end{align}
\end{subequations}
where $\tau_k = (2k - 1)\tau_{\text{gate}}/4$ denotes the pulse center, $\tau_R$ and $\tau_D$ control the temporal widths of the Rabi frequency and detuning, respectively, $\Delta_0$ is the detuning amplitude, and the normalization constants $C_\Omega = \sqrt{2e}$ and $C_\Delta = e$ ensure that the peak values reach $\Omega_{\text{max}}$ and $\Delta_0$, respectively. We optimize the single-pulse complete population inversion (CPI) in Ref.~\cite{Rangelov2010} into an echo pulse composed of two completely identical symmetric pulses. During this process, the dynamical phases accumulated by the two pulses are exactly canceled, leaving only the geometric phase $\pi$.

We combine zero-area pulses with a level-crossing-free architecture, endowing our scheme with distinctive physical attributes. The Rabi frequency in Eq.~\eqref{eq:Omega} is a time-odd function, satisfying $\Omega_k(\tau_k + t) = -\Omega_k(\tau_k - t)$; its time integral over a single pulse vanishes, forming a zero-area pulse. This zero-area property provides an intrinsic first-order insensitivity to slow intensity noise. The nonadiabatic transition probability scales as
\begin{equation}
	P_{\mathrm{nonad}}\sim|\int_{-\infty}^{+\infty}g(t)\,dt|^{2}.
\end{equation} Under a uniform rescaling fluctuation $\Omega(t)\to(1+\varepsilon)\Omega(t)$, the first-order correction to $g(t)$ can be written as a total time derivative,
\begin{equation}
	\delta g(t)=\frac{\varepsilon}{2}\frac{d}{dt}\left[\frac{\Omega(t)\Delta(t)}{\Omega^{2}(t)+\Delta^{2}(t)}\right]+\mathcal{O}(\varepsilon^{2}),
\end{equation}
whose integral over the full evolution interval vanishes because the boundary terms are equal for our even detuning and odd Rabi frequency (as shown in Supplementary Material III). Therefore, the first-order correction to the nonadiabatic transition probability vanishes:
\begin{equation}
	\delta P_{\mathrm{nonad}}=\mathcal{O}(\varepsilon^{2}).
\end{equation}
The detuning in Eq.~\eqref{eq:Delta} is a time-even function, satisfying $\Delta_k(\tau_k + t) = \Delta_k(\tau_k - t)$. It remains strictly positive throughout the evolution, approaching the resonance condition most closely at the pulse center ($\tau = \tau_k$) without crossing zero.

The pulse architecture shown in Fig.~\ref{fig:2}(a) determines the system's evolution trajectory in Bloch space to be fundamentally different from traditional RAP. Traditional RAP schemes require the laser frequency to chirp through the resonance point. At this point, the instantaneous eigenenergy gap is minimal, making the adiabatic condition most vulnerable to violation, leading to nonadiabatic transitions and limiting the achievable gate fidelity. In contrast, the level-crossing-free scheme adopted here maintains a strictly positive detuning. By avoiding level crossing and staying away from the region of minimal energy gap during the entire evolution, nonadiabatic leakage is effectively suppressed.

To elucidate the physical origin of the suppressed nonadiabatic leakage, we examine the nonadiabatic coupling
\begin{equation}
	g(t)=\frac{\langle-|\dot{H}|+\rangle}{\tilde{\Omega}(t)}=\frac{\dot{\Omega}\Delta-\Omega\dot{\Delta}}{2(\Omega^{2}+\Delta^{2})},
\end{equation}
where
\begin{equation}
	\tilde{\Omega}(t)=\sqrt{\Omega^{2}(t)+\Delta^{2}(t)}
\end{equation}is the instantaneous eigenenergy gap. Performing a Taylor expansion around the pulse center $t=0$ gives $\Omega(t)\approx At$ and $\Delta(t)\approx Bt^{2}$. Consequently, the nonadiabatic coupling approaches a finite constant
\begin{equation}
	g(t)\xrightarrow{t\to 0}-\frac{B}{2A},
\end{equation}
and the energy gap opens linearly as $\tilde{\Omega}(t)\sim A|t|$.

The mixing angle 
\begin{equation}
	\theta(t)=\arctan[\Omega(t)/\Delta(t)]
\end{equation}characterizes the rotation of the instantaneous eigenstates. Its time derivative
\begin{equation}
	\dot{\theta}(t)=\frac{\dot{\Omega}\Delta-\Omega\dot{\Delta}}{\Omega^{2}+\Delta^{2}},
\end{equation}
\begin{figure}[htbp]
	\includegraphics[width=0.5\textwidth]{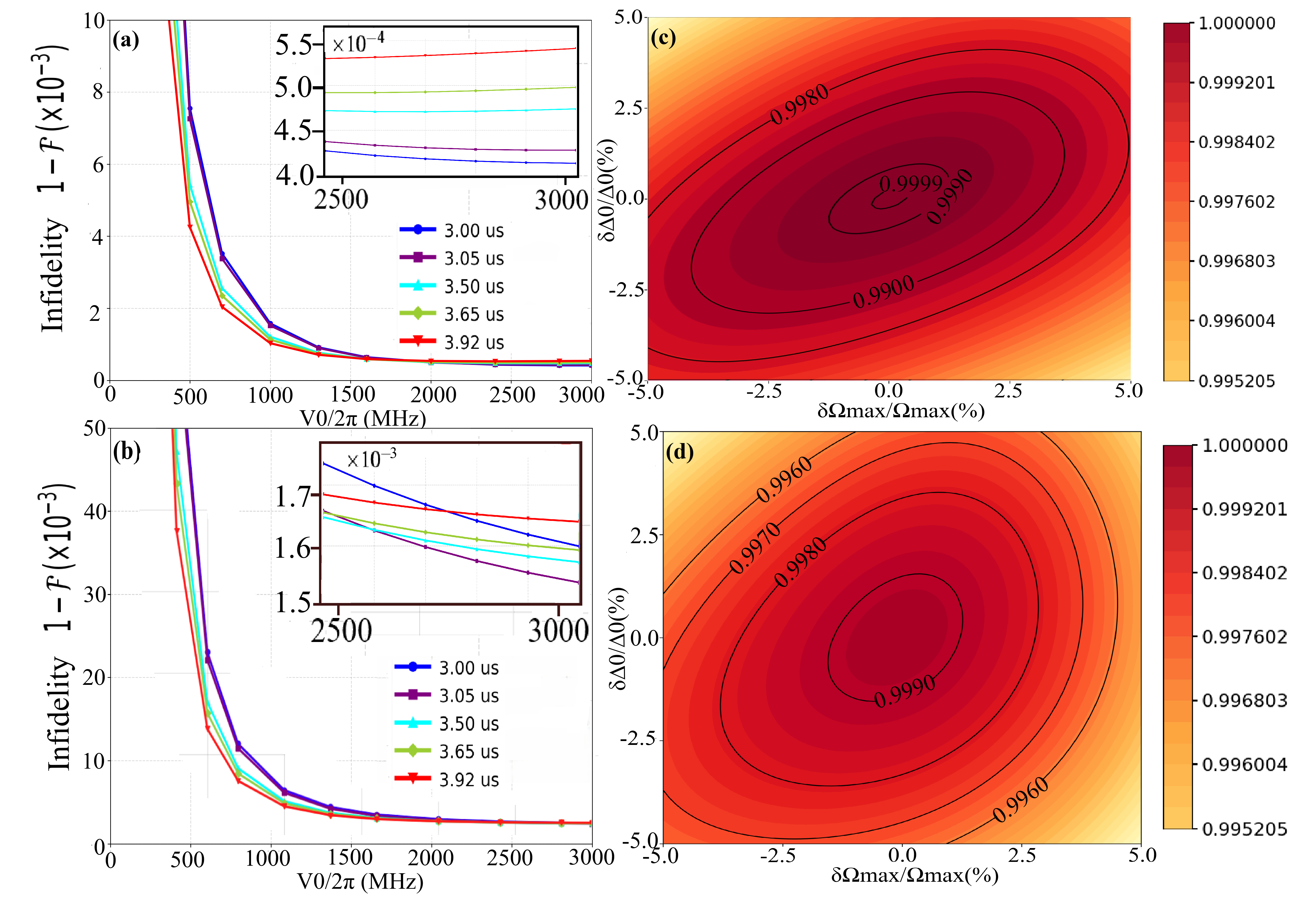}
	\captionsetup{justification=raggedright,singlelinecheck=false}
	\caption{Fidelity and parameter robustness analysis for CZ and CCZ gates. (a) and (b): Variation of the infidelity with $V_0$ at different total gate times for the CZ and CCZ gates, respectively. (c) and (d): Contour plots of the fidelity for the CZ and CCZ gates against parameter fluctuations, with $\Omega_{\text{max}} = (2\pi)\,14.93~\mathrm{MHz}$, $\Delta_0 = 0.45\,\Omega_{\text{max}}$, $\tau_R = 11.1$, $\tau_D = 8.58$ for the CZ gate, and $\Omega_{\text{max}} = (2\pi)\,13.75~\mathrm{MHz}$, $\Delta_0 = 0.14\,\Omega_{\text{max}}$, $\tau_R = 12$, $\tau_D = 9.15$ for the CCZ gate. The total gate time is $3.92\,\mu\text{s}$ for both gates.}
	\label{fig:3}
\end{figure}
tends to a finite value $-B/A$ as $t\to0$ (as shown in Supplementary Material II). This indicates that the rate of change of the adiabatic basis does not diverge at the resonance point, allowing the Bloch vector to follow the torque vector smoothly without nonadiabatic transitions.

Unlike conventional RAP, where $g(t)$ diverges at the level-crossing point, our level-crossing-free design keeps $g(t)$ finite. The minimal-gap region is compressed to $|t|\lesssim A/B$, and the standard adiabaticity ratio $|\dot{\tilde{\Omega}}/\tilde{\Omega}^{2}|$ decays as $1/|t|^{2}$ outside this region (as shown in Supplementary Material I and II). By choosing $A/B\ll\Omega_{\max}^{-1}$, the time spent traversing this narrow window is much shorter than the characteristic time scale of nonadiabatic transitions, thus preventing significant nonadiabatic losses.

Figure.~\ref{fig:2}(b) shows the instantaneous eigenenergy spectra of $H(t)$, including the two-atom and single-atom spectra. From the evolution of the instantaneous eigenenergy spectrum, the difference between the level-crossing-free scheme and traditional RAP is even more profound. In traditional echo RAP schemes~\cite{Xue2024}, because the detuning crosses the resonance point, the adiabatic energy spectra of two atoms and single atoms form an avoided crossing at the resonance moment, the instantaneous eigenenergy gap reaches its minimum value, and the adiabatic condition faces the greatest challenge; in the scheme of this paper, the detuning always remains positive, the adiabatic energy spectrum always maintains a larger energy gap spacing, effectively protecting the adiabatic condition. The Rydberg blockade effect decouples the doubly excited state $|rr\rangle$ from the single-excitation coupling subspace, maintaining the symmetry of the adiabatic energy spectrum, so that the dynamical phases accumulated by the two pulses can be precisely canceled. This energy spectrum symmetry ensures precise cancellation of dynamical phases across the two pulses, enabling high-fidelity geometric quantum gates within the level-crossing-free framework.

We employ two identical pulses of the form given in Eqs.~\eqref{eq:pulses} to construct a symmetric double-pulse sequence, thereby implementing the CZ gate, with a total gate time of $\tau_{\text{gate}}$. The specific evolution processes for each input computational basis under the gate operation are described as follows: for the input state $|00\rangle$, both atoms are in the dark state $|0\rangle$ and are unaffected by the laser driving, remaining unchanged as $|00\rangle$ throughout the gate operation. 
For the input state $|01\rangle$ (and $|10\rangle$), a single atom is in the $|1\rangle$ state and coupled to the Rydberg state $|r\rangle$. The system adiabatically evolves along the single-atom instantaneous eigenstate to $|0r\rangle$ (or $|r0\rangle$) during the first RAP pulse, and the second RAP pulse drives the system back to the respective initial state $|01\rangle$ or $|10\rangle$. The two pulses are completely symmetric, and the accumulated dynamical phases are precisely canceled due to time-reversal symmetry, leaving only the geometric phase $\pi$. For the input state $|11\rangle$, under the strong Rydberg blockade limit, the doubly excited state $|rr\rangle$ is effectively decoupled.

The system evolution is restricted to the two-dimensional subspace spanned by $|11\rangle$ and the symmetric single-excitation superposition $(|1r\rangle + |r1\rangle)/\sqrt{2}$, as shown in Fig.~\ref{fig:2}(c).The first RAP pulse adiabatically transfers $|11\rangle$ to $(|1r\rangle + |r1\rangle)/\sqrt{2}$, and the second RAP pulse fully restores the population to $|11\rangle$. The dynamical phase in the two-atom case is also precisely canceled between the two pulses, ultimately yielding the geometric phase $\pi$. In summary, after the two symmetric pulses, all computational bases achieve complete population return, and all except $|00\rangle$ accumulate the geometric phase $\pi$, thereby obtaining the matrix representation of the CZ gate, $U_{CZ} = \text{diag}(1, -1, -1, -1)$, thus realizing the CZ gate. The above logic can be directly extended to more control qubits. Taking the CCZ gate as an example, place three atoms at the vertices of an equilateral triangle, ensuring that each pair of atoms satisfies the blockade condition. The design of the three-qubit CCZ gate follows a similar approach to the CZ gate, with the Rabi frequency and detuning finely tuned. The three atoms are arranged at the vertices of an equilateral triangle to ensure symmetry under the Rydberg blockade mechanism between each pair of atoms~\cite{Evered2023, Tang2022}; the strong Rydberg interaction between adjacent atoms is a necessary condition for dynamical phase cancellation. The logical transformations of each input computational basis can be specifically described as: for the input state $|111\rangle$, the system adiabatically evolves to the symmetric single-excitation superposition $(|r11\rangle + |1r1\rangle + |11r\rangle)/\sqrt{3}$ during the first RAP pulse, and the second RAP pulse fully restores the population to $|111\rangle$ and accumulates the geometric phase, ultimately obtaining a negative sign. The states $|011\rangle$, $|101\rangle$, and $|110\rangle$ have only two atoms in the pulse-coupled state, and their evolution is exactly the same as the $|11\rangle$ case in the two-atom CZ gate. The states $|001\rangle$, $|010\rangle$, and $|100\rangle$ with only one atom in $|1\rangle$ evolve similarly to the single-atom case, completing full atomic population return. The state $|000\rangle$ is a dark state for the pulse and does not participate in evolution. After two symmetric RAP pulses, all computational basis states except $|000\rangle$ effectively cancel the dynamical phase and accumulate a geometric phase $\pi$, yielding the three-qubit CCZ gate matrix: $U_{CCZ} = \text{diag}(1, -1, -1, -1, -1, -1, -1, -1)$. Parameterized symmetric CZ gates~\cite{Li2022} and time-optimal Rydberg gate analyses~\cite{Jandura2022, Saffman2020, Mao2023, Wei2022} provide useful comparisons for the present level-crossing-free design.

We verify the feasibility of the proposed scheme via numerical integration of the time-dependent Schr\"odinger equation. The optimized pulse parameters are chosen as $\Omega_{\text{max}} = (2\pi)\,9.7~\mathrm{MHz}$, $\Delta_0 = 0.60\,\Omega_{\text{max}}$, $\tau_R = 12$, $\tau_D = 17$, and the total gate time $\tau_{\text{gate}} = 240$, with the Rydberg interaction strength set to $V_0 = 167\,\Omega_{\text{max}}$. Figure.~\ref{fig:2}(c) shows the population evolution of the relevant states for each computational basis. When initially in $|11\rangle$, the population is transferred to $(|r1\rangle + |1r\rangle)/\sqrt{2}$ during the first pulse, with transfer efficiency close to 1; the second RAP pulse completes the full population return. The population of $|rr\rangle$ is always suppressed below $10^{-4}$ throughout the process, indicating that the Rydberg blockade effect is effectively maintained. The single-atom evolution is also shown as a dashed line in Fig.~\ref{fig:2}(c): the $|1\rangle$ population first decreases and then returns to 1, showing a corresponding single-peak structure, clearly demonstrating the complete atomic population return process. The inset shows the Bloch sphere trajectory of the two-atom system in the blockade limit, evolving in the effective two-level subspace: with $|11\rangle$ and $(|r1\rangle + |1r\rangle)/\sqrt{2}$ as effective basis vectors, the state vector starts from $|11\rangle$, passes near the equator, and returns, completing a closed loop symmetric about the sphere center. The geometric phase is half of the solid angle subtended by the Bloch-sphere trajectory, which equals $\pi$ for this evolution.

We define the gate fidelity as the overlap between the actual evolved final state and the output of the ideal CZ (or CCZ) gate. To comprehensively evaluate the performance, we choose the symmetric superposition of all computational bases as the initial state, making the fidelity sensitive to various types of errors.

The CZ gate infidelity $1 - F$ as a function of $V_0/2\pi$ under different total gate times is shown in Fig.~\ref{fig:3}(a). When $V_0$ is small, incomplete blockade leads to a slight population in $|rr\rangle$ and imperfect cancellation of the dynamical phase, resulting in higher infidelity. As $V_0$ increases, the curve rapidly decreases and saturates beyond $V_0/2\pi = 1500$~MHz. At a gate time of $t_{\text{gate}} = 3.92~\mu\text{s}$, the minimum infidelity is approximately $10^{-4}$, corresponding to a fidelity of $F = 0.9999$. A comparison of different gate times reveals a trade-off: longer gate times favor the adiabatic condition but increase dissipation accumulation, whereas shorter gate times have the opposite effect.

Figure.~\ref{fig:3}(b) shows the corresponding results for the CCZ gate. Since the three atoms share collective excitation, the effective Rabi frequency scales as $\sqrt{3}$, requiring a slightly higher blockade strength compared to the CZ gate. At a gate time of $t_{\text{gate}} = 3.92~\mu\text{s}$, the minimum infidelity is approximately $10^{-3}$, corresponding to a fidelity of $F = 0.999$. The curve also exhibits a saturation trend with increasing $V_0$, indicating that the scheme retains good feasibility in multi-qubit extensions. Recent high-fidelity Rydberg gate benchmarks report CZ fidelities up to 0.997--0.999 with detailed error budgeting~\cite{Tsai2025, Motzoi2009}, supporting the physical plausibility of our parameter choices.

We simulate open-system dynamics within the framework of the Lindblad master equation to investigate the influence of dissipation:
\begin{equation}\label{eq:lindblad}
\partial_t \hat{\rho}(t) = -i[\hat{H}(t), \hat{\rho}] + \sum_k \mathcal{D}_k[\hat{\rho}],
\end{equation}
where the dissipators are defined as
\begin{equation}
\mathcal{D}_k[\hat{\rho}] = \hat{L}_k \hat{\rho} \hat{L}_k^\dagger - \frac{1}{2}\left(\hat{L}_k^\dagger \hat{L}_k \hat{\rho} + \hat{\rho} \hat{L}_k^\dagger \hat{L}_k\right),
\end{equation}
and $\{\hat{L}_k\}$ are the jump operators describing the dissipation channels. For $^{133}\mathrm{Cs}$ atoms, we employ the following operators: $\hat{L}_1 = \sqrt{\gamma_r/16}\,|0\rangle\langle r|$ and $\hat{L}_2 = \sqrt{\gamma_r/16}\,|1\rangle\langle r|$ to model spontaneous decay into the ground-state manifold, and $\hat{L}_3 = \sqrt{7\gamma_r/8}\,|r\rangle\langle r|$ to account for dephasing. The total decay rate is set to $\gamma_{r}=1/(540~\mu\mathrm{s})$~\cite{Sibalic2017}.

The CZ gate is shown in Fig.~\ref{fig:3}(c), where the optimal fidelity is located at the center of the parameter plane, approximately $0.9999$. The fidelity contours reveal anisotropy: the fidelity decays more rapidly along the detuning axis than along the Rabi frequency axis, showing higher sensitivity to detuning offsets than to Rabi amplitude fluctuations. Specifically, when the Rabi frequency fluctuation is controlled within $\pm 2\%$, and the detuning offset is within $\pm 1\%$, the fidelity remains above $0.999$. This anisotropic robustness originates from the first-order cancellation of low-frequency intensity noise by the zero-area pulse, while adiabatic evolution inherently tolerates slow detuning drifts.

The CCZ gate is shown in Fig.~\ref{fig:3}(d), where the central optimal fidelity is approximately 0.999. The 0.995 contour covers most of the parameter plane, only slightly below this value in the four corners (joint fluctuations). Under single-parameter fluctuations of $\pm 3\%$, the fidelity is always higher than 0.997. Compared with Ref.~\cite{Xue2024}, our scheme outperforms the original protocol in terms of the central fidelity for both the CZ and CCZ gates. It is worth noting that the overall tolerance of the CCZ gate to errors is comparable to that of the CZ gate, and even superior in some regions, which stems from the statistical averaging effect of multi-bit collective excitation on individual parameter deviations.

Our numerical results can be deeply understood from the physical protection mechanism of the level-crossing-free scheme. The obvious anisotropy of the CZ gate fidelity contours directly stems from the first-order noise suppression effect of zero-area pulses: when the Rabi frequency undergoes overall scaling, because the pulse area is always zero, the first-order response of the system is completely canceled, and the fidelity decrease is mainly contributed by second-order and higher-order effects, so the decay is slower. In contrast, detuning offsets change the contact condition at the pulse center, but this change's effect on the adiabatic evolution trajectory is adiabatic---the system still evolves in the same half-space; only the extremum of the mixing angle is slightly offset, so the impact on fidelity is relatively small. For the CCZ gate, its robustness is even better than that of the CZ gate in some regions. This counterintuitive phenomenon can be understood from the statistical averaging effect of collective excitation: the local parameter deviation of any one of the three atoms' influence on the collective excited state is diluted to $1/3$ of the total, and deviations between different atoms may partially cancel each other, thereby improving the overall gate operation's tolerance to individual noise. This collective robustness is another important advantage of the level-crossing-free scheme in multi-qubit extensions.

In summary, we propose level-crossing-free echoing rapid adiabatic passage pulses with global laser dressing to implement high-fidelity controlled-Z gates with Rydberg atoms. The level-crossing-free design avoids the minimal eigenenergy gap region by maintaining a strictly positive detuning that approaches the resonance condition most closely only at the pulse center, suppressing nonadiabatic transitions. In conventional echoing RAP schemes, the detuning must sweep through the resonance point, causing an avoided crossing in the adiabatic energy spectra and exposing the system to nonadiabatic leakage. In contrast, our level-crossing-free scheme maintains a larger energy gap spacing throughout the entire evolution, satisfying the adiabatic condition and delivering a CZ gate fidelity of $0.9999$.

The zero-area Rabi pulse suppresses first-order laser intensity noise: the vanishing time integral of the Rabi pulse cancels the first-order response of the system to low-frequency intensity fluctuations, and numerical simulations show that the scheme is more sensitive to detuning offsets than to intensity noise. The two-segment symmetric structure cancels the dynamical phase via time-reversal symmetry, removing the need for additional phase compensation. All driving fields smoothly ramp down at the end of the gate, and global laser addressing removes the requirement for rapid spatially resolved switching. The scheme is also applicable to other quantum platforms including superconducting circuits, ion traps, and solid-state spin systems~\cite{DelGrosso2025, Ying2023, Losey2024, Philbin2021, Zhang2023}, as well as to other Rydberg species such as Rb, Sr, and Yb, and supports parallel execution of multi-pair qubit gates in large-scale arrays.~\cite{Poole2025, Borregaard2020}

See the Supplementary Material for a detailed description of the pulse optimization, the adiabaticity analysis, the zero-area pulse properties, the fidelity definition, and the open-system Lindblad dynamics for both CZ and CCZ gates.

This work was supported by the National Natural Science Foundation of China (Nos.~61505100 and 12574375), the Fundamental Research Program of Shanxi Province (Grant No.~202203021211301), and the Research Project Supported by Shanxi Scholarship Council of China (Nos.~2023-028 and 2022-014).

\bibliographystyle{apsrev4-1}
\bibliography{refs}

\newpage
\pagebreak
\widetext
\begin{center}
	\textbf{\large Supplemental Materials: Robust controlled-Z gate for Rydberg atoms based on level-crossing-free echoing rapid adiabatic passage}
\end{center}

\setcounter{equation}{0}
\setcounter{figure}{0}
\setcounter{table}{0}
\setcounter{section}{0}
\setcounter{page}{1}
\makeatletter
\renewcommand{\theequation}{S\arabic{equation}}
\renewcommand{\thefigure}{S\arabic{figure}}
\renewcommand{\bibnumfmt}[1]{[S#1]}
\renewcommand{\thesection}{\arabic{section}}
\renewcommand{\thesubsection}{\arabic{section}.\arabic{subsection}}
\section{THEORETICAL SCHEME}
The no-level-crossing RAP pulse we propose for implementing the controlled-Z gate is optimized based on the pulse scheme for achieving complete population inversion (CPI), as shown in Fig.~\ref{fig:S1}. The detuning and Rabi frequency of this original pulse are given by
\begin{equation}
	\Delta(t)=\Delta_0 \frac{t^2}{T^2}\left[\operatorname{sech}\left(\frac{t+\tau}{T}\right)+\operatorname{sech}\left(\frac{t-\tau}{T}\right)\right],
\end{equation}
	
\begin{equation}
	\Omega(t)=\Omega_0 \frac{t}{T} e^{-(t/2.5T)^2},
\end{equation}
where $\Delta(t)$ is even-symmetric: $\Delta(-t)=\Delta(t)$, and satisfies $\Delta(t)\ge 0$ for all $t$, vanishing only at $t=0$; $\Omega(t)$ is odd-symmetric: $\Omega(-t)=-\Omega(t)$, is a zero-area pulse, and satisfies $\Omega(0)=0$ with opposite signs on the two sides. The two-level Hamiltonian is
\begin{equation}
	H(t)=\frac{\hbar}{2}
	\begin{bmatrix}
		-\Delta(t) & \Omega(t) \\
		\Omega(t) & \Delta(t)
	\end{bmatrix},
\end{equation}
with the instantaneous eigenstates
\begin{equation}
	|+\rangle = \begin{pmatrix} \cos(\theta/2) \\ \sin(\theta/2) \end{pmatrix},\qquad
	|-\rangle = \begin{pmatrix} -\sin(\theta/2) \\ \cos(\theta/2) \end{pmatrix},
	\label{eq:eigenstates}
\end{equation}
and the instantaneous eigenenergy gap
\begin{equation}
	\tilde{\Omega}(t)=\sqrt{\Omega^2(t)+\Delta^2(t)}.
	\label{eq:gap}
\end{equation}

We perform a Taylor expansion of the pulse forms for the level-crossing-free CPI around $t=0$:
\begin{equation}
	\Omega(t)=\frac{\Omega_0}{T}t+\mathcal{O}(t^3),
	\label{eq:6}
\end{equation}
\begin{equation}
	\Delta(t)=2\Delta_0\,\text{sech}(\tau/T)\frac{t^2}{T^2}+\mathcal{O}(t^4),
	\label{eq:7}
\end{equation}
where $A=\Omega_0/T$ and $B=2\Delta_0\,\text{sech}(\tau/T)/T^2$, then we have
\begin{equation}
	\Omega(t)=At+\mathcal{O}(t^3),
	\label{eq:8}
\end{equation}
\begin{equation}
	\Delta(t)=Bt^2+\mathcal{O}(t^4).
	\label{eq:9}
\end{equation}
The time derivatives are
\begin{equation}
	\dot{\Omega}(t)=A+\mathcal{O}(t^2),\qquad \dot{\Delta}(t)=2Bt+\mathcal{O}(t^3).
	\label{eq:10}
\end{equation}
The instantaneous energy gap is
\begin{equation}
	\tilde{\Omega}(t)=\sqrt{A^2t^2+B^2t^4+\mathcal{O}(t^6)}
	=|t|\sqrt{A^2+B^2t^2+\mathcal{O}(t^4)}.
	\label{eq:11}
\end{equation}
This shows that $\tilde{\Omega}(t)\to 0$ as $t\to 0$, meaning that the instantaneous energy gap closes at the pulse center. However, for sufficiently small but nonzero $|t|$, the energy gap opens linearly with $|t|$. Consequently, the region where the gap remains extremely small is confined to a very narrow neighborhood of $t=0$, namely $|t|\lesssim A/B$.

To quantitatively assess the validity of the adiabatic evolution under the present pulse configuration, we examine the standard adiabatic condition:
\begin{figure}[H]
	\centering
	\includegraphics[width=1\linewidth]{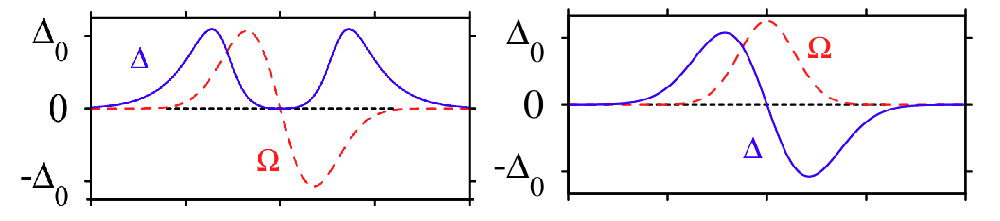}
	\caption{The left panel shows the pulse diagram for the CPI without level crossing, while the right panel shows that with level crossing.}
	\label{fig:S1}
\end{figure}
\begin{equation}
	\left| \frac{\dot{\tilde{\Omega}}(t)}{\tilde{\Omega}^2(t)} \right| \ll 1, 
\end{equation}
which quantifies the rate of change of the instantaneous eigenenergy gap relative to the square of the gap itself, serving as a necessary criterion for the adiabatic approximation to hold.

From Eq.~\eqref{eq:11}, we obtain

\begin{equation}
	\tilde{\Omega}^2(t) = A^2 t^2 + B^2 t^4 + \mathcal{O}(t^6),
\end{equation}
Differentiating Eq.~\eqref{eq:11} with respect to time, we obtain the time 
derivative of $\tilde{\Omega}(t)$:

\begin{equation}
	\dot{\tilde{\Omega}}(t) = \frac{2A^2 t + 4B^2 t^3 + \mathcal{O}(t^5)}
	{2\sqrt{A^2 t^2 + B^2 t^4 + \mathcal{O}(t^6)}}.
\end{equation}
Consequently, the adiabaticity ratio is given by

\begin{equation}
	\left| \frac{\dot{\tilde{\Omega}}(t)}{\tilde{\Omega}^2(t)} \right|
	= \frac{|A^2 + 2B^2 t^2 + \mathcal{O}(t^4)|}
	{|t|^2 (A^2 + B^2 t^2 + \mathcal{O}(t^4) )^{3/2}}.
\end{equation}
For $|t| \gg A/B$, this ratio decays as $1/|t|^2$, and the adiabatic condition is well satisfied. Within the minimal-gap region $|t| \lesssim A/B$, although the ratio increases, the system spends only an extremely short time in this narrow window, whose width scales as $A/B$. By choosing the pulse parameters such that $A/B \ll \Omega_{\max}$, the time spent traversing this region is much shorter than the characteristic time scale of nonadiabatic transitions, thus preventing significant nonadiabatic losses.

This analysis stands in sharp contrast to conventional RAP schemes, where the minimum gap occurs near the detuning crossing point, and the system must traverse the entire resonance region. In conventional RAP, the system remains in the ``dangerous region'' for a longer time, making it difficult to satisfy the adiabatic condition throughout the entire evolution. In contrast, the level-crossing-free design of our scheme compresses the minimal-gap region to a narrow neighborhood of a single time point, effectively alleviating the burden of the adiabatic condition.
 \begin{figure}[H]
 	\centering
 	\includegraphics[width=0.9\linewidth]{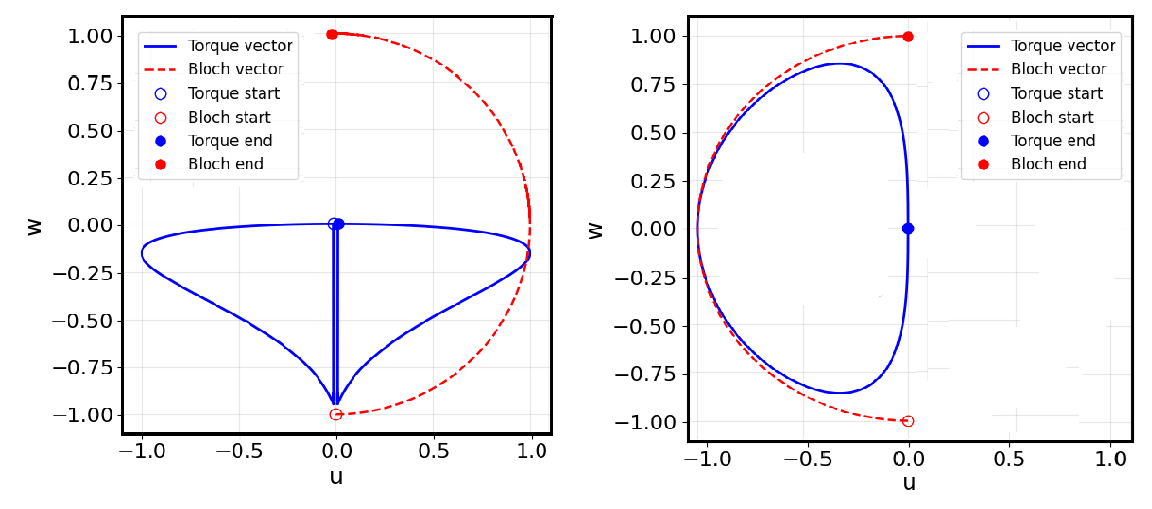}
 	\caption{Motion of the Bloch vector (dashed red line) and the normalized torque vector (solid blue line) in the $(u,w)$ plane. The left panel shows the trajectory for the level-crossing-free CPI, which corresponds to the scheme adopted in the present work. The right panel shows the Bloch-vector trajectory for the conventional level-crossing CPI. The hollow and filled circles mark the initial and final positions of the Bloch vector, respectively.}
 	\label{fig:S2}
 \end{figure}
 
\begin{figure}[H]
	\centering
	\includegraphics[width=0.9\linewidth]{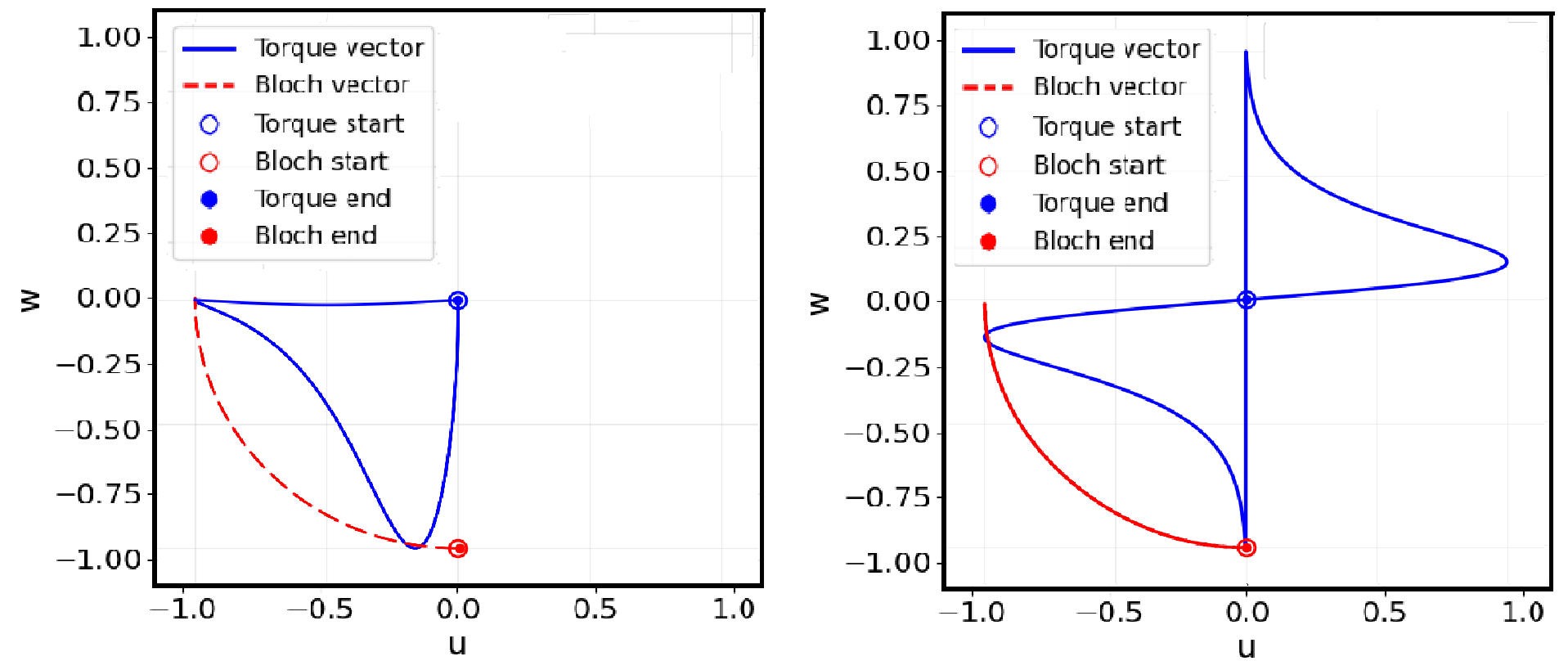}
	\caption{Motion of the Bloch vector (dashed red line) and the normalized torque vector (solid blue line) in the $(u,w)$ plane. The left panel shows the trajectory for the level-crossing-free CPR, where both the Rabi frequency and the detuning are even functions of time. The right panel shows the Bloch-vector trajectory for the conventional level-crossing CPR, where both the Rabi frequency and the detuning are odd functions of time. The hollow and filled circles mark the initial and final positions of the Bloch vector, respectively.}
	\label{fig:S3}
\end{figure}

To provide an intuitive geometric picture of the level-crossing-free adiabatic population transfer adopted in this work, the Bloch-vector trajectories shown in Fig.~\ref{fig:S2} contrast the conventional level-crossing CPI scenario with the level-crossing-free CPI structure; the Bloch-vector trajectories shown in Fig.~\ref{fig:S3} contrast the conventional level-crossing complete population return(CPR) scenario with the level-crossing-free CPR structure. 

For the level-crossing-free CPI, the Rabi frequency is an odd function of time, while the detuning is an even function and remains non-negative, so the diabatic energy curves touch but never cross. As shown in the left panel of Fig.~\ref{fig:S2}, the Bloch vector starts from $w=-1$ and follows the torque vector to $w=+1$, achieving CPI without ever traversing a level-crossing point. This is the single-atom building block of our echoing double-pulse sequence: the first pulse drives the Bloch vector from the south pole to the north pole, and the second pulse (identical and time-reversed) drives it back, accumulating a geometric phase $\pi$ while the dynamical phase is canceled by the time-reversal symmetry.

The conventional CPI with level crossing features an even Rabi frequency and an odd detuning, so the diabatic energy curves cross at $t=0$. As shown in the right panel of Fig.~\ref{fig:S2}, the Bloch vector starts from $w=-1$ (state $|1\rangle$) and adiabatically follows the torque vector to $w=+1$ (state $|r\rangle$), realizing complete population inversion. The trajectory passes through the equator ($w=0$) at the level-crossing point, where the instantaneous energy gap is minimal, and the adiabatic condition is most vulnerable to violation.

For the level-crossing-free CPR, both the Rabi frequency and the detuning are even functions of time. The detuning touches resonance at $t=0$ but never crosses it, while the Rabi frequency remains positive and symmetric. The mixing angle $\theta(t)$ rotates from $0$ to $\pi/2$ and then back to $0$. As shown in the left panel of Fig.~\ref{fig:S3}, the Bloch vector starts from $w=-1$, rises to the equator at $t=0$ (where the torque vector is oriented along the $u$-axis), and then descends back to the south pole $w=-1$, resulting in complete population return. Since both the coupling and the detuning maintain their respective signs throughout the entire process, the torque vector always stays within the same quadrant of the $(u,w)$ plane, merely executing a "turnaround" that drives the Bloch vector along a closed trajectory.

The CPR with level crossing features both the Rabi frequency and the detuning as odd functions of time. The diabatic energy curves still cross, but the Rabi frequency additionally changes sign at $t=0$, corresponding to a zero-area pulse. The mixing angle $\theta(t)$ rotates from $0$ to $\pi/2$ and then back to $0$. As shown in the right panel of Fig.~\ref{fig:S3}, the Bloch vector starts from $w=-1$, and as the torque vector swings toward the upper half-plane, it temporarily deviates from the south pole but eventually returns to $w=-1$, because the asymptotic mixing angle recovers its initial value. This complete population return originates from the overall antisymmetry of the Hamiltonian and holds not only in the adiabatic limit but also exactly in the general case.

Among the cases presented above, only the level-crossing-free CPI and the conventional level-crossing CPI achieve complete population inversion, where the Bloch vector starts from the south pole ($w=-1$) and ends at the north pole ($w=+1$). In contrast, the two CPR cases return the Bloch vector to the south pole, resulting in no net population transfer.

For the implementation of a controlled-Z gate, CPI is the essential building block because it creates the required conditional phase accumulation: the $\ket{11}$ blockade subspace must be transferred to the Rydberg state and back with a geometric phase $\pi$, while the single-excitation subspaces ($\ket{01}$ and $\ket{10}$) remain unaffected by the blockade. This conditional dynamics relies critically on the population ending in the Rydberg state at the end of the first pulse, rather than returning to the ground state as in CPR.

Moreover, the level-crossing-free CPI is preferred over the level-crossing CPI because the detuning never crosses zero, eliminating the extended dangerous region near the resonance where the adiabatic gap is small and nonadiabatic transitions are most likely to occur. This feature makes the level-crossing-free CPI particularly robust against parameter fluctuations and thus more suitable for high-fidelity quantum gate operations.

\section{NONADIABATIC COUPLING}

The mixing angle is defined as the parameter that characterizes the mixing of the instantaneous eigenstates in the bare-state basis, given by
\begin{equation}
	\theta(t)=\arctan\frac{\Omega(t)}{\Delta(t)},
\end{equation}
substituting Eqs.~\eqref{eq:8} and \eqref{eq:9} yields
\begin{equation}
	\theta(t)=\arctan\left[\frac{At+\mathcal{O}(t^3)}{Bt^2+\mathcal{O}(t^4)}\right]
	=\arctan\left[\frac{A}{B}\frac{1}{t}+\mathcal{O}(t)\right].
\end{equation}

Hence, $\theta\to +\pi/2$ as $t\to 0^+$ and $\theta\to -\pi/2$ as $t\to 0^-$, indicating that the mixing angle undergoes a $\pi$ jump from $-\pi/2$ to $+\pi/2$ at $t=0$.

The time derivative of $\theta(t)$ is given by
\begin{equation}
	\dot{\theta}(t)=\frac{d}{dt}\arctan\frac{\Omega}{\Delta}
	=\frac{\dot{\Omega}\Delta-\Omega\dot{\Delta}}{\Omega^2+\Delta^2},
\end{equation}
substituting Eqs.~\eqref{eq:8}, \eqref{eq:9}, and \eqref{eq:10}, the numerator and denominator are respectively
\begin{equation}
	\dot{\Omega}\Delta-\Omega\dot{\Delta}
	=(A)(Bt^2)-(At)(2Bt)+\mathcal{O}(t^4)
	=-ABt^2+\mathcal{O}(t^4),\label{eq:19}
\end{equation}
\begin{equation}
	\Omega^2+\Delta^2=A^2t^2+B^2t^4+\mathcal{O}(t^4)
	=A^2t^2\left[1+\frac{B^2}{A^2}t^2+\mathcal{O}(t^2)\right].\label{eq:20}
\end{equation}
Therefore,
\begin{equation}
	\dot{\theta}(t)=\frac{-ABt^2+\mathcal{O}(t^4)}{A^2t^2+B^2t^4+\mathcal{O}(t^5)}
	=-\frac{B}{A}+\mathcal{O}(t^2).
\end{equation}
This shows that $\dot{\theta}(t)$ tends to the finite value -$B/A$ as $t\to 0$, indicating that the rate of change of the mixing angle does not diverge.

The nonadiabatic coupling term is defined as
\begin{equation}
	g(t)=\frac{\langle -|\dot H|+\rangle}{\tilde{\Omega}(t)}
	=\frac{\dot{\Omega}\Delta-\Omega\dot{\Delta}}{2(\Omega^2+\Delta^2)},
    \label{eq:22}
\end{equation}
substituting Eq.~\eqref{eq:19}, and Eq.~\eqref{eq:20} into Eq.~\eqref{eq:22}, we obtain
\begin{equation}
	g(t)=\frac{-ABt^2+\mathcal{O}(t^4)}{2\left[A^2t^2+B^2t^4+\mathcal{O}(t^4)\right]}
	=\frac{-ABt^2+\mathcal{O}(t^4)}{2A^2t^2\left[1+\frac{B^2}{A^2}t^2+\mathcal{O}(t^2)\right]}.
\end{equation}

As $t\to 0$,
\begin{equation}
	g(t)\sim \frac{-ABt^2}{2A^2t^2}
	=-\frac{B}{2A}.
\end{equation}

Unlike conventional RAP, where the nonadiabatic coupling typically diverges near the resonance point, in our scheme the nonadiabatic coupling $g(t)$ remains finite as $t \to 0$, approaching the constant value $-B/(2A)$. The instantaneous energy gap $\tilde{\Omega}(t)$ itself closes linearly with $|t|$, but the system does not undergo an actual nonadiabatic transition. Nonadiabatic transitions are most common when the detuning changes sign, corresponding to a level crossing. In our scheme, $\Delta(t) \ge 0$ for all $t$, so the two levels only touch without crossing. At $t = 0$, the system naturally resides in one of the instantaneous eigenstates. The mixing angle $\theta(t)$ undergoes a $\pi$ jump at $t = 0$, but the system only passes through the resonance point without crossing it. This level-touching design ensures that the system spends only a very short time in the minimal-gap region, leaving insufficient time to accumulate nonadiabatic transition probability, thereby eliminating the dynamical source of nonadiabatic transitions.

In conventional RAP, the Rabi frequency $\Omega(t)$ is usually chosen as an even function of time, while the detuning $\Delta(t)$ crosses zero at $t = 0$ and changes sign. In the vicinity of $\Delta = 0$, the instantaneous eigenenergy gap $\tilde{\Omega}(t) = \sqrt{\Omega^2(t) + \Delta^2(t)}$ attains its minimum value, making the adiabatic condition most stringent. As a result, the system unavoidably traverses a finite-width ``dangerous region'' around the resonance point, where nonadiabatic couplings remain significant. To suppress these nonadiabatic transitions, a sufficiently large peak Rabi frequency $\Omega_0$ is required, which inevitably introduces extra experimental imperfections.

\section{ZERO-AREA PULSES}

Each constituent pulse is designed such that the Rabi frequency is an odd function of time with respect to its center $\tau_k$,
\begin{equation}
	\Omega_k(\tau_k + t) = -\Omega_k(\tau_k - t),
\end{equation}
which ensures that the pulse area vanishes identically,
\begin{equation}
	\mathcal{A}_k = \int_{-\infty}^{+\infty} \Omega_k(\tau) \, d\tau = 0.\label{eq:26}
\end{equation}
This zero-area property provides an intrinsic first-order insensitivity to slow intensity noise, as demonstrated below.

Consider a small uniform rescaling fluctuation $\varepsilon$ ($|\varepsilon|\ll 1$) of the Rabi frequency:
\begin{equation}
	\Omega(t) \to \Omega'(t) = (1+\varepsilon)\Omega(t).
\end{equation}
\begin{figure}[H]
	\centering
	\includegraphics[width=0.6\linewidth]{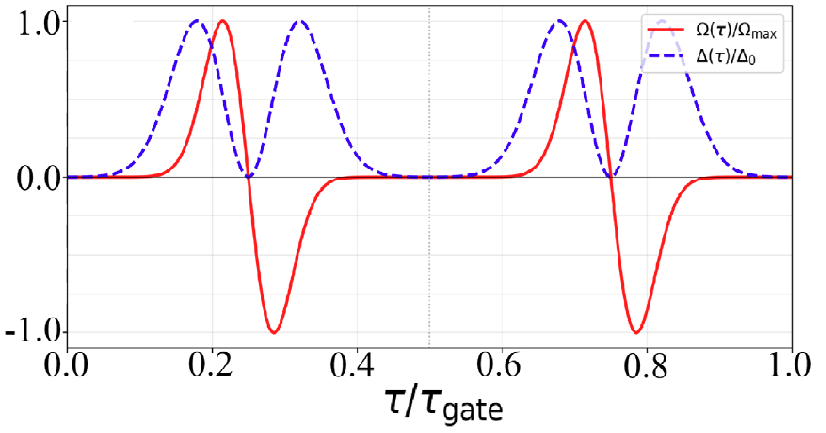}
	\caption{Two identical level-crossing
		free pulses. The Rabi frequency $\Omega(t)$ crosses zero and changes sign at the pulse 
		center, while the detuning $\Delta(t)$ does not cross the level-crossing 
		point.}
	\label{fig:S4}	
\end{figure}
The first-order change in the pulse area is
\begin{equation}
	\delta \mathcal{A} = \int_{-\infty}^{+\infty} \Omega'(t)\,dt - \int_{-\infty}^{+\infty} \Omega(t)\,dt
	= \varepsilon \int_{-\infty}^{+\infty} \Omega(t)\,dt.
\end{equation}
By the zero-area condition Eq.~\eqref{eq:26}, $\int_{-\infty}^{+\infty}\Omega(t)\,dt = 0$, hence
\begin{equation}
	\delta \mathcal{A} = 0.
\end{equation}
That is, a uniform rescaling of the Rabi frequency does not alter the pulse area to first order.

The nonadiabatic transition probability is governed by the integral of the nonadiabatic coupling term $g(t)$ over the evolution interval:
\begin{equation}
	P_{\mathrm{nonad}} \sim \left| \int_{-\infty}^{+\infty} g(t) \, dt \right|^2.
\end{equation}
When the Rabi frequency is subject to a uniform rescaling fluctuation $\Omega(t) \to (1+\varepsilon)\Omega(t)$, corresponding to first-order relative intensity noise, the first-order correction to $g(t)$ is
\begin{equation}
	\delta g(t) = \varepsilon \left( \Omega \frac{\partial g}{\partial \Omega} + \dot{\Omega} \frac{\partial g}{\partial \dot{\Omega}} \right) + O(\varepsilon^2).
\end{equation}
From Eq.~\eqref{eq:22}, we obtain
\begin{equation}
	\frac{\partial g}{\partial \Omega}
	= \frac{-\dot{\Delta}(\Omega^2+\Delta^2) - 2\Omega(\dot{\Omega}\Delta-\Omega\dot{\Delta})}
	{2(\Omega^2+\Delta^2)^{2}},
\end{equation}
\begin{equation}
	\frac{\partial g}{\partial \dot{\Omega}} = \frac{\Delta}{2(\Omega^2 + \Delta^2)}.
\end{equation}
Thus, we get
\begin{equation}
	 \Omega \frac{\partial g}{\partial \Omega} + \dot{\Omega} \frac{\partial g}{\partial \dot{\Omega}}
	= \frac{-\dot{\Delta}\Omega(\Omega^2+\Delta^2) - 2\Omega^2(\dot{\Omega}\Delta-\Omega\dot{\Delta})+\dot{\Omega}\Delta(\Omega^2+\Delta^2)}
	{2(\Omega^2+\Delta^2)^{2}}.
\end{equation}
On the other hand, taking the time derivative of $\frac{\Omega\Delta}{\Omega^2+\Delta^2}$ gives

\begin{equation}
	\frac{d}{dt}\left[\frac{\Omega\Delta}{\Omega^2+\Delta^2}\right]
	= \frac{(\dot{\Omega}\Delta+\Omega\dot{\Delta})(\Omega^2+\Delta^2) - \Omega\Delta(2\Omega\dot{\Omega}+2\Delta\dot{\Delta})}
	{(\Omega^2+\Delta^2)^2}.
\end{equation}
Expanding and simplifying the numerator, one can verify that

\begin{equation}
	\Omega \frac{\partial g}{\partial \Omega} + \dot{\Omega} \frac{\partial g}{\partial \dot{\Omega}}
	= \frac{1}{2}\frac{d}{dt}\left[\frac{\Omega\Delta}{\Omega^2+\Delta^2}\right].
\end{equation}
Its first-order term can be written as a total time derivative:
\begin{equation}
	\delta g(t)  = \frac{\varepsilon}{2} \frac{d}{dt} \left[ \frac{\Omega(t)\Delta(t)}{\Omega^2(t)+\Delta^2(t)} \right] + \mathcal{O}(\varepsilon^2).
\end{equation}
Integrating $\delta g(t)$ over the full time interval yields
\begin{equation}
	\int_{-\infty}^{+\infty} \delta g(t) \, dt
	= \frac{\varepsilon}{2} \left[ \frac{\Omega(t)\Delta(t)}{\Omega^2(t)+\Delta^2(t)} \right]_{-\infty}^{+\infty}
	+ \mathcal{O}(\varepsilon^2).
\end{equation}
For the zero-area pulses employed in our scheme, where the detuning $\Delta(t)$ is an even function of time, the two boundary terms are exactly equal, so their difference vanishes. Therefore, the first-order correction to the nonadiabatic transition probability vanishes:
\begin{equation}
	\delta P_{\mathrm{nonad}} = \mathcal{O}(\varepsilon^2).
\end{equation}

\section{LEVEL-CROSSING-FREE RAP PULSES AND FIDELITY}
The pulse shapes in our scheme are given by
\begin{align}
	\Omega_k(\tau) &= -\Omega_{\max} C_\Omega \frac{\tau-\tau_k}{\tau_R}
	\exp\!\left[-\left(\frac{\tau-\tau_k}{\tau_R}\right)^2\right], \\
	\Delta_k(\tau) &= \Delta_0 C_\Delta \left(\frac{\tau-\tau_k}{\tau_D}\right)^2
	\exp\!\left[-\left(\frac{\tau-\tau_k}{\tau_D}\right)^2\right].
\end{align}
To determine the normalization constants, we maximize the envelope functions. For $\Omega_k$, let $x=(\tau-\tau_k)/\tau_R$, then the envelope of $\Omega_k$ is $f(x)=x e^{-x^2}$,
whose maximum satisfies
\begin{equation}
	f'(x)=e^{-x^2}(1-2x^2)=0,
\end{equation}
yielding $x=1/\sqrt{2}$, and thus
\begin{equation}
	f_{\max}=\frac{1}{\sqrt{2e}}.
\end{equation}
From the normalization condition $C_\Omega f_{\max}=1$, we obtain
\begin{equation}
	C_\Omega=\sqrt{2e}.
\end{equation}
Similarly, let $y=(\tau-\tau_k)/\tau_D$, then the envelope of $\Delta_k$ is
$g(y)=y^2 e^{-y^2}$. From
\begin{equation}
	g'(y)=2y e^{-y^2}(1-y^2)=0,
\end{equation}
the maximum occurs at $y=1$, giving $g_{\max}=1/e$. From the normalization
condition $C_\Delta g_{\max}=1$, we obtain
\begin{equation}
	C_\Delta=e.
\end{equation}
Thus far, the normalization constants for the pulse parameters $\Omega_k(\tau)$ and $\Delta_k(\tau)$ have been completely specified. The factor $C_\Omega = \sqrt{2e}$ guarantees that the Rabi frequency reaches its peak value $\Omega_{\max}$, while $C_\Delta = e$ ensures that the detuning attains its maximum $\Delta_0$. These normalization conditions are essential for the precise realization of the pulse waveforms in the numerical simulations that follow.

We now define the fidelity that will be used to assess the gate performance.

In a closed system, given an initial pure state $|\psi\rangle$, the actual evolution yields the pure state $|\psi_f\rangle = U_{\rm actual}|\psi\rangle$, while the ideal final state is the pure state $|\psi_{\rm id}\rangle = U_{\rm id}|\psi\rangle$. The fidelity between two pure states is naturally defined as the overlap probability:

\begin{equation}
	\mathcal{F}(|\psi_f\rangle, |\psi_{\rm id}\rangle) = |\langle\psi_{\rm id}|\psi_f\rangle|^2. \label{eq:38}
\end{equation}

In terms of density matrices, $\rho_f = |\psi_f\rangle\langle\psi_f|$ and $\rho_{\rm id} = |\psi_{\rm id}\rangle\langle\psi_{\rm id}|$, Eq.~\eqref{eq:38} can be rewritten as

\begin{equation}
	\mathcal{F}(\rho_f, \rho_{\rm id}) = \mathrm{Tr}[\rho_f \rho_{\rm id}],
\end{equation}

since
\begin{align}
	\mathrm{Tr}[\rho_f \rho_{\rm id}]
	&= \mathrm{Tr}[|\psi_f\rangle\langle\psi_f| \, |\psi_{\rm id}\rangle\langle\psi_{\rm id}|] \notag \\
	&= |\langle\psi_f|\psi_{\rm id}\rangle|^2 \notag \\
	&= |\langle\psi_{\rm id}|\psi_f\rangle|^2.
\end{align}
Thus, in the pure-state case, the fidelity equals the trace of the product of the two density matrices, which is equivalently the expectation value of the projector onto the ideal state with respect to the actual state:

\begin{equation}
	\mathcal{F}(\rho_f, \rho_{\rm id}) = \langle\psi_{\rm id}|\rho_f|\psi_{\rm id}\rangle. \label{eq:41}
\end{equation}
When the system is subject to dissipation, the final state after evolution is a mixed-state density matrix $\rho_f$, while the ideal final state remains a pure state $|\psi_{\rm id}\rangle$. In this case, directly replacing $\rho_f$ in Eq.~\eqref{eq:41} with the mixed-state density matrix yields the fidelity between the mixed state and the pure state:

\begin{equation}
	\mathcal{F} = \langle \psi_{\rm id} | \rho_f | \psi_{\rm id} \rangle. \label{eq:42}
\end{equation}
This expression strictly reduces to Eq.~\eqref{eq:38} in the pure-state limit $\rho_f = |\psi_f\rangle\langle\psi_f|$. Substituting the initial probe state $|\psi\rangle$ and the ideal gate operator $U_{\rm id}$ into Eq.~\eqref{eq:42}, we obtain

\begin{equation}
	\mathcal{F} = \langle \psi | U_{\rm id}^{\dagger} \rho_f U_{\rm id} | \psi \rangle. 
\end{equation}
For the CZ gate, $U_{\rm id} = U_{CZ}$,
where $|\psi\rangle$ is a fixed symmetric probe state,
is the symmetric-input process fidelity.

To quantify the gate performance, we employ the symmetric superposition of all computational basis states as the initial probe state,
\begin{equation}
	|\psi_{\mathrm{sym}}\rangle=\frac{1}{\sqrt{d}}\sum_{k=0}^{d-1}|k\rangle,\qquad d=2^{n},
\end{equation}
where $n$ is the number of qubits. For the two-qubit CZ gate ($d=4$), this reads explicitly
\begin{equation}
	|\psi_{\mathrm{sym}}\rangle=\frac{1}{2}\bigl(|00\rangle+|01\rangle+|10\rangle+|11\rangle\bigr).
\end{equation}

The ideal CZ gate is diagonal in the computational basis, $U_{\mathrm{id}}=\mathrm{diag}(1,-1,-1,-1)$, so the ideal output state is
\begin{equation}
	|\psi_{\mathrm{id}}\rangle=U_{\mathrm{id}}|\psi_{\mathrm{sym}}\rangle=\frac{1}{2}\bigl(|00\rangle-|01\rangle-|10\rangle-|11\rangle\bigr).
\end{equation}

\section{OPEN SYSTEM DYNAMICS AND LINDBLAD MASTER EQUATION}
To investigate the influence of dissipative effects on the gate fidelity, we establish the Lindblad master equation for the two-atom system within the single-excitation subspace. Because the Rydberg blockade effect excludes the doubly excited state $|rr\rangle$ from the evolution subspace and $|00\rangle$ remains a dark state unaffected by the laser, the entire gate operation can be strictly restricted to low-dimensional subspaces, significantly reducing the numerical cost.

The full two-atom space is spanned by the computational basis $\{|00\rangle,|01\rangle,|10\rangle,|11\rangle,|0r\rangle,|r0\rangle,|1r\rangle,|r1\rangle,|rr\rangle\}$. In the interaction picture, the evolution of the density matrix $\rho(t)$ satisfies
\begin{equation}
	\partial_{t}\rho(t)=-i[\hat{H}(t),\rho]+\mathcal{D}_{k}[\rho],
\end{equation}
where the dissipative superoperators are defined as
\begin{equation}
	\mathcal{D}_{k}[\rho]=\hat{L}_{k}\rho\hat{L}_{k}^{\dagger}-\frac{1}{2}\left(\hat{L}_{k}^{\dagger}\hat{L}_{k}\rho+\rho\hat{L}_{k}^{\dagger}\hat{L}_{k}\right).
\end{equation}
For $^{133}\mathrm{Cs}$ atoms, we employ the three jump operators stated in the main text:
\begin{align}
	\hat{L}_{1}&=\sqrt{\frac{\gamma_{r}}{16}}\,|0\rangle\langle r|\otimes\hat{I}_{2},\\[2pt]
	\hat{L}_{2}&=\sqrt{\frac{\gamma_{r}}{16}}\,|1\rangle\langle r|\otimes\hat{I}_{2},\\[2pt]
	\hat{L}_{3}&=\sqrt{\frac{7\gamma_{r}}{8}}\,|r\rangle\langle r|\otimes\hat{I}_{2},
\end{align}
where $\gamma_{r}=1/(540\,\mu\mathrm{s})$ is the total radiative decay rate of the Rydberg state $|r\rangle$. $\hat{L}_{1}$ and $\hat{L}_{2}$ describe spontaneous decay from $|r\rangle$ into the two ground hyperfine levels $|0\rangle$ and $|1\rangle$ each with a branching ratio of $1/16$, while $\hat{L}_{3}$ describes pure dephasing of $|r\rangle$ accounting for the remaining $7/8$ branching ratio. Because both atoms are driven simultaneously by a global laser, the above operators also act symmetrically on the second atom, giving six jump operators in total.

To demonstrate the accumulation of gate errors and the post-pulse relaxation behavior, we plot the time-dependent infidelity $1-\mathcal{F}(t)$ for both the CZ and CCZ gates within the dissipative Lindblad dynamics framework. For the CZ gate, we employ, for each atom, two jump operators describing the spontaneous decay from $\ket{r}$ to the ground hyperfine levels $\ket{0}$ and $\ket{1}$, respectively, and one pure-dephasing operator acting on states containing $\ket{r}$, giving six jump operators in total. The CCZ gate uses the corresponding three-atom generalization. The initial state is the symmetric superposition of all computational basis states, $|\psi_{\mathrm{sym}}\rangle$, given by Eq.~(S52) for the CZ gate and its three-qubit extension for the CCZ gate. The time-dependent fidelity is computed as $\mathcal{F}(t)=\langle\psi_{\mathrm{id}}|\rho(t)|\psi_{\mathrm{id}}\rangle$, where the ideal target state is $|\psi_{\mathrm{id}}\rangle=U_{\mathrm{id}}|\psi_{\mathrm{sym}}\rangle$. The pulse sequence consists of two identical level-crossing-free pulses with a total gate time of $\tau_{\mathrm{gate}}=3.00\,\mu\mathrm{s}$.

For the CZ gate, the parameters are $\Omega_{\max}/2\pi = 40\,\text{MHz}$, $\Delta_0 = 6.60\,\text{MHz}$, $\tau_R = 0.50\,\mu\text{s}$, $\tau_D = 0.22\,\mu\text{s}$, and $V_0/2\pi = 200\,\text{MHz}$, while for the CCZ gate, the parameters are $\Omega_{\max}/2\pi = 32\,\text{MHz}$, $\Delta_0 = 10.0\,\text{MHz}$, $\tau_R = 0.18\,\mu\text{s}$, $\tau_D = 0.27\,\mu\text{s}$, and $V_0/2\pi = 100\,\text{MHz}$. With these parameters, we now examine the origin of the oscillatory behavior observed in the infidelity curves.

The laser driving induces periodic population transfer between $|1\rangle$ and $|r\rangle$. Although the pulse is designed for adiabatic evolution, the system always remains in a superposition of instantaneous eigenstates of the time-dependent Hamiltonian rather than being strictly locked to a single adiabatic branch. Different computational basis components experience different effective couplings: the single-atom excitation subspace has an effective Rabi frequency $\Omega(t)$; the two-atom blockade subspace ($|11\rangle$) acquires a collective enhancement to $\sqrt{2}\,\Omega(t)$; and the three-atom CCZ gate has $\sqrt{3}\,\Omega(t)$. These Rabi oscillations at distinct frequencies are encoded in the density matrix $\rho(t)$, while the fidelity $\mathcal{F}(t)=\langle\psi_{\mathrm{id}}|\rho(t)|\psi_{\mathrm{id}}\rangle$ measures the overlap between $\rho(t)$ and a fixed target projector. As $\rho(t)$ rotates periodically in Hilbert space, this overlap necessarily oscillates.

The initial state is the symmetric superposition $|\psi_{\mathrm{sym}}\rangle$ of all computational bases, and each component follows a distinct path: $|00\rangle$ remains a dark state; $|01\rangle$ and $|10\rangle$ undergo single-atom two-level oscillations; $|11\rangle$ undergoes two-atom blockade oscillations; and $|111\rangle$ undergoes three-body collective oscillations. The total fidelity is an incoherent weighted average of these contributions. The superposition of oscillations at different frequencies produces beats, leading to complex local maxima and minima in the overall curve.

\end{document}